\documentclass[a4paper,fleqn,usenatbib]{mnras}

\usepackage{mathptmx}
\usepackage[T1]{fontenc}
\usepackage{ae,aecompl}

\usepackage{graphicx}	% Including figure files
\usepackage{amsmath}	% Advanced maths commands
\usepackage{amssymb}	% Extra maths symbols
\usepackage{gensymb}	% Extra maths symbols 
\usepackage{natbib} %Biblography 
\title[Analysis of horizontal flows]{Analysis of horizontal flows in the solar granulation}

\author[C. Quintero Noda et al.]{C. Quintero Noda,$^{1}$\thanks{E-mail: carlos@solar.isas.jaxa.jp}
T. Shimizu,$^{1}$
Y. Suematsu,$^{2}$
\\
$^{1}$Institute of Space and Astronautical Science, Japan Aerospace Exploration Agency, Sagamihara, Kanagawa 252-5210, Japan\\
$^{2}$National Astronomical Observatory of Japan, 2-21-1 Osawa, Mitaka, Tokyo 181-8588, Japan\\
}

\date{Accepted 2016 January 8. Received 2016 January 7; in original form 2015 October 23}
\pubyear{2015}

\begin{document}
\label{firstpage}
\pagerange{\pageref{firstpage}--\pageref{lastpage}}
\maketitle

% Abstract of the paper
\begin{abstract}
Solar limb observations sometimes reveal the presence of a satellite lobe in the blue wing of the Stokes~$I$ profile from pixels belonging to granules. The presence of this satellite lobe has been associated in the past to strong line of sight gradients and, as the line of sight component is almost parallel to the solar surface, to horizontal granular flows. We aim to increase the knowledge about these horizontal flows studying a spectropolarimetric observation of the north solar pole. We will make use of two state of the art techniques, the spatial deconvolution procedure that increases the quality of the data removing the stray light contamination, and spectropolarimetric inversions that will provide the vertical stratification of the atmospheric physical parameters where the observed spectral lines form. We inverted the Stokes profiles using a two component configuration, obtaining that one component is strongly blueshifted and displays a temperature enhancement at upper photospheric layers while the second component has low redshifted velocities and it is cool at upper layers. In addition, we examined a large number of cases located at different heliocentric angles, finding smaller velocities as we move from the centre to the edge of the granule. Moreover, the height location of the enhancement on the temperature stratification of the blueshifted component also evolves with the spatial location on the granule being positioned on lower heights as we move to the periphery of the granular structure.  
\end{abstract}

% Select between one and six entries from the list of approved keywords.
% Don't make up new ones.
\begin{keywords}
Sun: photosphere -- Sun: granulation -- methods: data analysis
\end{keywords}

%%%%%%%%%%%%%%%%%%%%%%%%%%%%%%%%%%%%%%%%%%%%%%%%%%

%%%%%%%%%%%%%%%%% BODY OF PAPER %%%%%%%%%%%%%%%%%%

\section{Introduction}

Solar plasma can harbour large velocities in the quiet solar photosphere. These high speed velocities are usually translated into characteristic spectral features. From highly dopplershifted circular polarization signals with complex profile shapes as single-lobed \citep{Borrero2010,QuinteroNoda2014b,Jafarzadeh2015} or multi-lobed profiles \citep{Shimizu2008,QuinteroNoda2014a} to secondary satellite lobes in the intensity profiles \citep{BellotRubio2009}. The reason behind this high speed plasma is still under debate and it seems that there is not a single answer. In fact, the studies performed on observations suggest magnetic reconnection \citep{Borrero2013,QuinteroNoda2014a}, convective collapse \citep{Nagata2008,Shimizu2008,Requerey2014} or siphon flows \citep{QuinteroNoda2014b} as possible trigger mechanisms. However, the mentioned mechanisms imply the presence of magnetic fields while in some cases, as the work of \cite{BellotRubio2009}, the characteristic spectral feature, a satellite lobe on the Stokes $I$ profile, seems to be related with the existence of non-magnetized horizontal flows. In that sense, a theoretical explanation can be found on \cite{Stein1998} where the authors said that the flow in granules resembles a flowing fountain where the fluid moves up inside the granule and then flows out toward and over its edge. Because of the small density scale height in the visible photosphere, the flow above the granule has to be strongly horizontally divergent in order to conserve mass, producing high speed flows toward the periphery of the granule. Observational spectral signatures of horizontal granular flows, from secondary satellite lobes in the Stokes $I$ profile, were unequivocally found for the first time on \citep{BellotRubio2009} using solar limb observations. We aim to continue the studies of the latter work analysing polarimetric observations of the north solar pole. The novelty of this paper is that we take advantage of two different techniques, the spatial deconvolution of spectropolarimetric data and the inversion of the Stokes profiles. Moreover, we intend to examine these events at different spatial locations on the granular structure found at different heliocentric angles. Therefore, we plan to comprehend the nature of the horizontal flows found on granular regions.

\begin{figure*}
\centering
\vspace{-0.9cm}
\includegraphics[width=16cm]{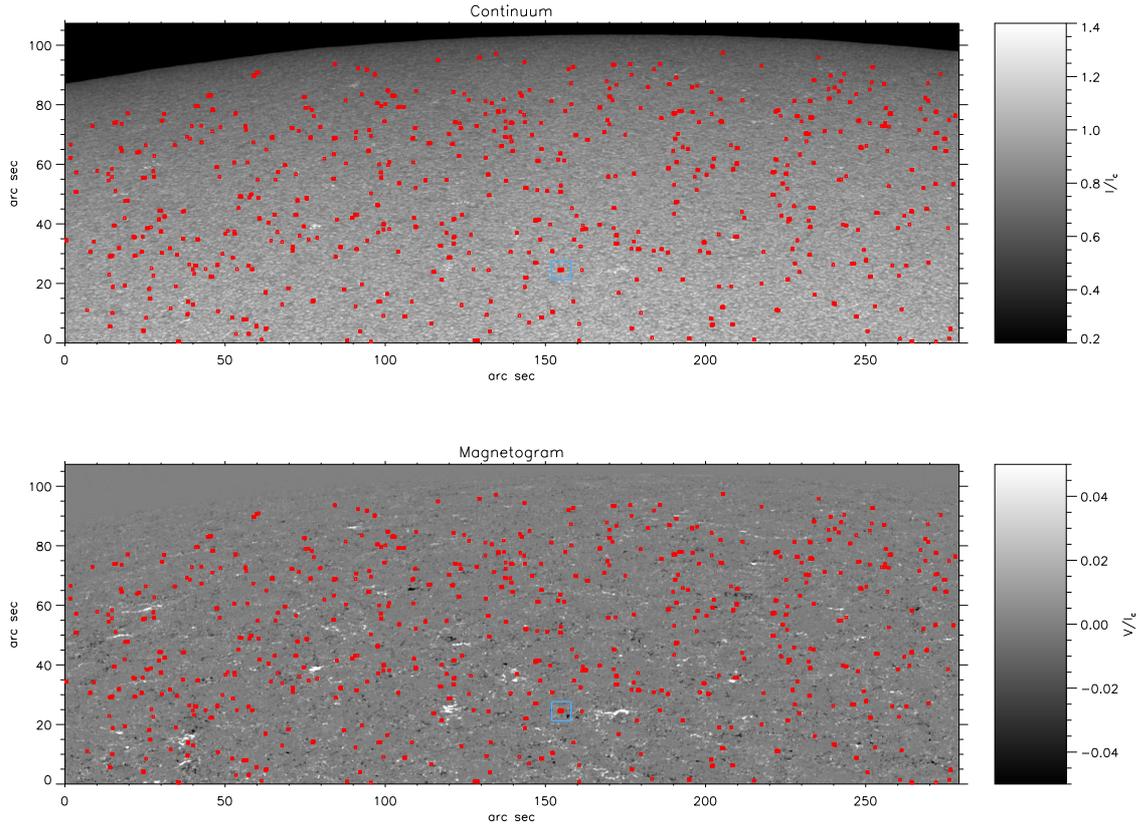}
\caption{Observed map of the north solar pole. Continuum signal (top) and magnetogram (bottom). Red colour indicates the position of pixels that show satellite lobes on the blue wing of Stokes $I$ profiles. The size of each coloured box is larger than the own event in order to facilitate its visualization.}
\label{cont}
\end{figure*}

\section{Data analysis}

\subsection{Observations and deconvolution}

The observation we used was taken by the Spectropolarimeter ($SP$) \citep{Lites2013} on board $Hinode/SOT$ \citep{Kosugi2007,Tsuneta2008,Suematsu2008,Shimizu2008SOT} on 2007 September $6^{\rm th}$ between 00:28-05:57 UT. $Hinode/SP$ recorded the Stokes $I$, $Q$, $U$, and $V$ profiles for the Fe~{\sc i} 6301.5 and 6302.5 \AA \ spectral lines. It used a spectral sampling of 21.55~m\AA \ and a spatial resolution of 0.32 arcsec. The integration time was 9.6 s per slit position what produces a noise value of approximately $1\times10^{-3}$ of the continuum signal ($I_c$) for the magnetic Stokes parameters \citep[for more information, see][]{Ichimoto2008,Lites2013SP_Prep}.

We  carried out the spatial deconvolution on the observed data following \cite{QuinteroNoda2015}. We chose a number of principal component analysis eigenvectors equal to (20, 7, 5, 10) for ($I$, $Q$, $U$, $V$). We performed 7 iterations in the deconvolution process. The original continuum contrast was 4.5 per cent and the value obtained after the deconvolution process is 7 per cent. We stopped the iteration process in that step because it provides an increase factor similar to the one obtained in previous works \citep{QuinteroNoda2015}. The changes induced by the deconvolution process on the observed polar map are described in detail on \cite{QuinteroNoda2016}, however, we want to stress that the deconvolution method slightly enhances the satellite lobe we intend to examine in this work.

\subsection{Detection of events}\label{criteria}

As pointed out by \cite{BellotRubio2009}, the satellite lobe is only present in the blue wing of Stokes $I$. Therefore, to find the position of pixels that show satellite lobes, we only examined the spectral region that covers the blue wing of Stokes $I$. We take as reference the Fe~{\sc i} 6301.5~\AA \ line and we compute the intensity difference between one spectral position and the following for the wavelength range $\Delta \lambda=[-260,-108]$~m\AA \ from the Fe~{\sc i} 6301.5~\AA \ line core. We establish as threshold, the following condition,

\begin{equation}\label{equ}
I(\Delta \lambda _i) \le I(\Delta \lambda _{i+1})*f_r
\end{equation}
where $i$ goes from zero to seven. For example, if $i=0$, $\Delta \lambda _0=-260$ m\AA \ and $\Delta \lambda _1=-237$ m\AA. The element $f_r$ is a reduction factor to avoid small spectral fluctuations, as the ones produced by the noise. However, in our case, we are examining deconvolved data and, therefore, the noise is almost negligible. Thus, we chose a value close to the unity, i.e. $f_r=0.98$. The Stokes $I$ profile, in normal conditions, displays larger intensity values as we move from the line core to continuum wavelengths. The condition applied in this work looks for cases where a monotonic increase does not happen for the spectral range of the blue wing. Thus, any pixel that satisfies Equation \ref{equ} for any index $i$ is considered in this work. Additionally, we include a second condition due to the presence of emission profiles when we examine off-limb pixels. We only take into account Stokes $I$ profiles that display a continuum intensity value larger than 0.6 $I_c$. We define $I_c$ as the mean continuum intensity obtained from the pixels of the bottom of the observed map, see Figure \ref{cont}, that are located at heliocentric values of $0.40<\mu<0.45$.

\section{Results}\label{profile}

Figure \ref{cont} shows in the top panel the continuum signal of the selected region from the observed map used in this work. We omitted part of the off-limb upper region from the original observation. Bottom row shows a Fe~{\sc i} 6302.5 \AA \ magnetogram, taken at $\pm$100 m\AA \ from the line centre. We marked in red the locations of Stokes $I$ profiles that display a second satellite lobe in the blue wing. Their presence is ubiquitous and it seems that they are not directly related neither with bright faculae, see upper panel, nor with magnetic field concentrations, bottom panel. Therefore, as pointed out by \cite{BellotRubio2009}, the physical mechanism behind the satellite lobes seems to be unrelated to any magnetic field configuration. This property indicates that they cannot be associated to the events studied in \citep{QuinteroNoda2014b}, discarding the possibility that these horizontal flows were the connection between the opposite footpoints analysed in the cited work.

We estimate the density of events taking into account how many pixels of the observed map fulfils the threshold criteria imposed, see Section \ref{criteria}. The results reveal that 0.12 per cent of total pixels shows a satellite lobe in the Stokes $I$ profile, indicating that this phenomenon is relatively common in the polar observations. Moreover, these results are in agreement with \cite{BellotRubio2009} who obtained a slightly larger value of 0.3 per cent. The difference could be due to a more demanding threshold from our side, although the results are still comparable. However, as was explained in the aforementioned work, these values are smaller than the ones predicted by the simulations \citep{Stein1998}, i.e. 3-4 per cent. 

\begin{figure}
%\hspace{+0.5cm}
\includegraphics[width=8.5cm]{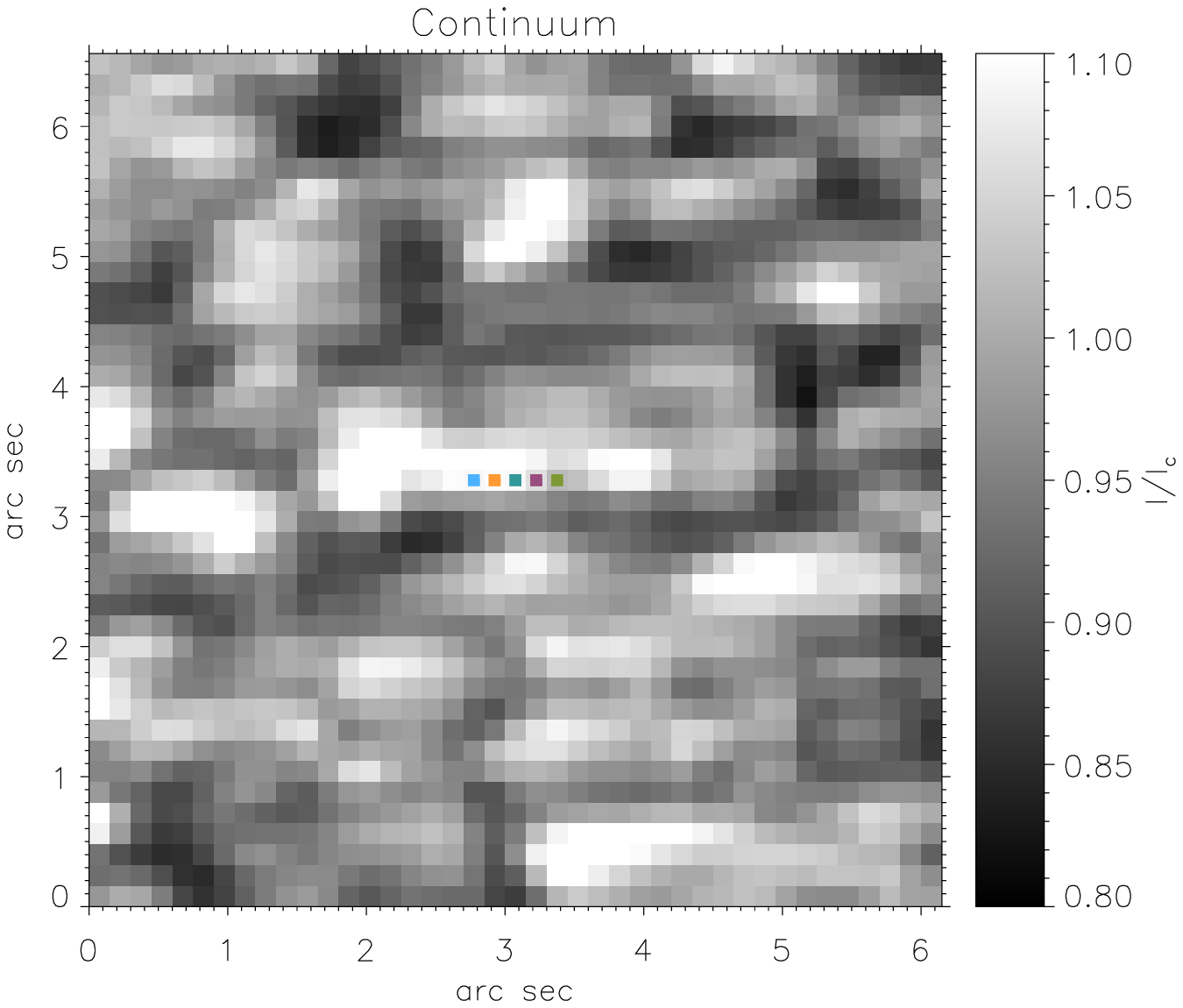}
\includegraphics[width=8.5cm]{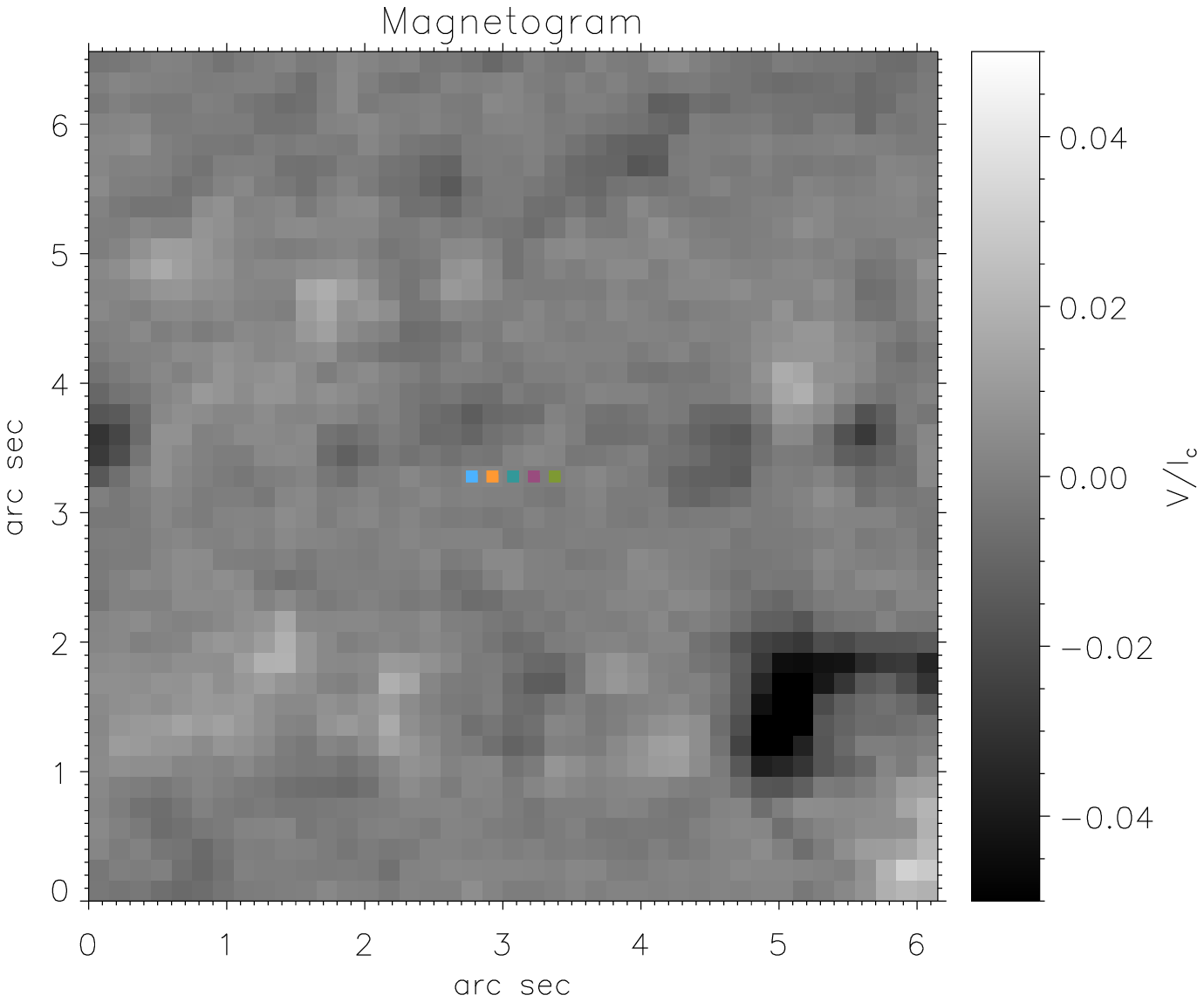}
\vspace{-0.4cm}
\caption{Reduced field of view centred in the example we examine in Section \ref{profile}, see also blue box on Figure \ref{cont}. We plot the continuum signal (top) and the magnetogram (bottom). Coloured squares designate the position of Stokes $I$ with satellite lobes in the blue wing.}
\label{cont_redux}
\end{figure} 

\begin{figure*}
%\centering
%\hspace{-0.9cm}
\includegraphics[width=17.0cm]{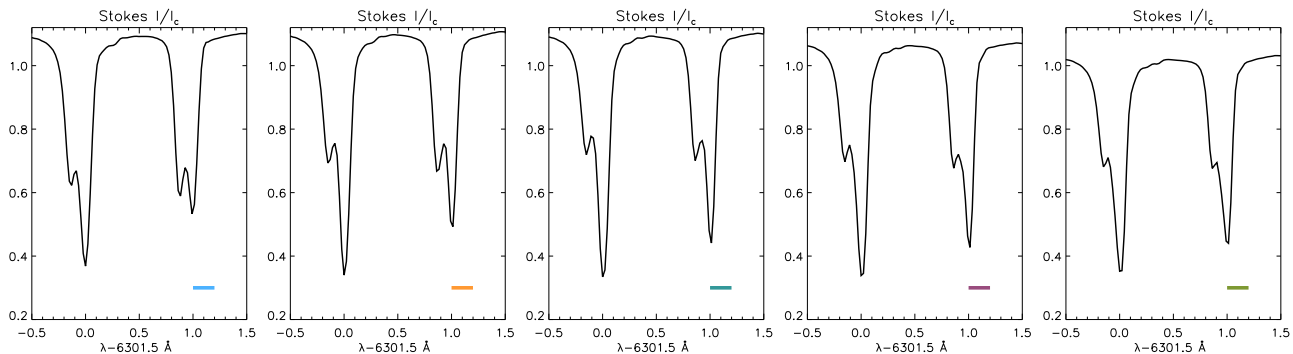}
\caption{Stokes $I$ profiles from the event displayed on Figure \ref{cont_redux}. We indicate with coloured lines at the bottom of each panel their spatial location on the mentioned figure.}
\label{pix}
\end{figure*}

As we want to perform a detailed analysis of the physical properties of these events, first we opted to select a single case to examine its characteristics, although we plan to make a statistical study in subsequent sections. This event is enclosed by the blue box of Figure \ref{cont}. We plotted in Figure \ref{cont_redux} a reduced field of view to see in detail the spatial distribution over the continuum (top) and magnetogram (bottom) maps. We marked with different colours the location of pixels that show satellite lobes. We can see, that they are positioned on top of a single granule from almost the centre to the edge of the structure. Regarding the magnetogram, we can see that the line of sight polarization signals are weak in this region.

We plot in Figure \ref{pix} the Stokes $I$ profiles marked with coloured squares on Figure \ref{cont_redux}. If we examine the continuum signal we can see that there is a decrease from the leftmost to the rightmost panel, coinciding with the spatial differences between the central part and the edge of the granule, see Figure \ref{cont_redux}. At the same time, although it is difficult to see by visual inspection, the satellite lobe, placed at the blue wing of both iron lines, is slightly moving from almost the core of the line (leftmost panel) to the blue wing of the line (rightmost panel). Therefore, we found a relation between the spatial location on the granule and the spectral position of the satellite lobe on the line. Additionally, the satellite lobe is usually more prominent in the Fe~{\sc i} 6302.5 \AA \ line and appears located at higher Stokes $I$ intensity values in the Fe~{\sc i} 6301.5 \AA \ line. We plan to deepen the details of the physical nature of the satellite lobe in the following section, performing inversions of the Stokes profiles. However, we want to mention before that we have not included in Figure \ref{pix} the polarization profiles because, generally, they are in the order of the noise level of the original data, thus, we cannot extract reliable information from them.

\section{Inversion of Stokes profiles}

We obtain the physical information of the atmospheric parameters inverting the Stokes profiles that display a satellite lobe in the blue wing of Stokes $I$. We carry out the inversion of the Stokes profiles using the SIR \citep[Stokes Inversion based on Response functions;][]{RuizCobo1992} code, which allows us to infer the optical depth dependence of these atmospheric parameters at each pixel independently.

\subsection{Configuration}\label{config}

As the deconvolution process removes the stray light contamination produced by the spatial point spread function of the telescope, we do not need to include a stray light component. However, the complexity of the Stokes $I$ profile demands the use of two different components. One component is going to be highly dopplershifted and will produce the satellite lobe while the other will generate the larger and almost non-shifted Stokes $I$ profile. On the other hand, polarimetric signals are most of the times at the noise level of the original data showing intensities below $3\sigma$, i.e. $3\times10^{-3}$ of $I_c$. In this manner, we cannot be sure about the reliability of the information we could infer from these low polarization signals. Therefore, we decided to focus in the inversion of the Stokes $I$ profiles and do not take into account the polarimetric profiles in the inversion. We consider that the satellite lobe is created by a non-magnetised material and we will not invert the magnetic field vector. We invert the gas temperature and the LOS plasma velocity, allowing gradients along the LOS in the temperature stratification while the LOS velocity is constant with height.

\begin{figure*}
\centering
%\hspace{-0.9cm}
\includegraphics[width=16.5cm]{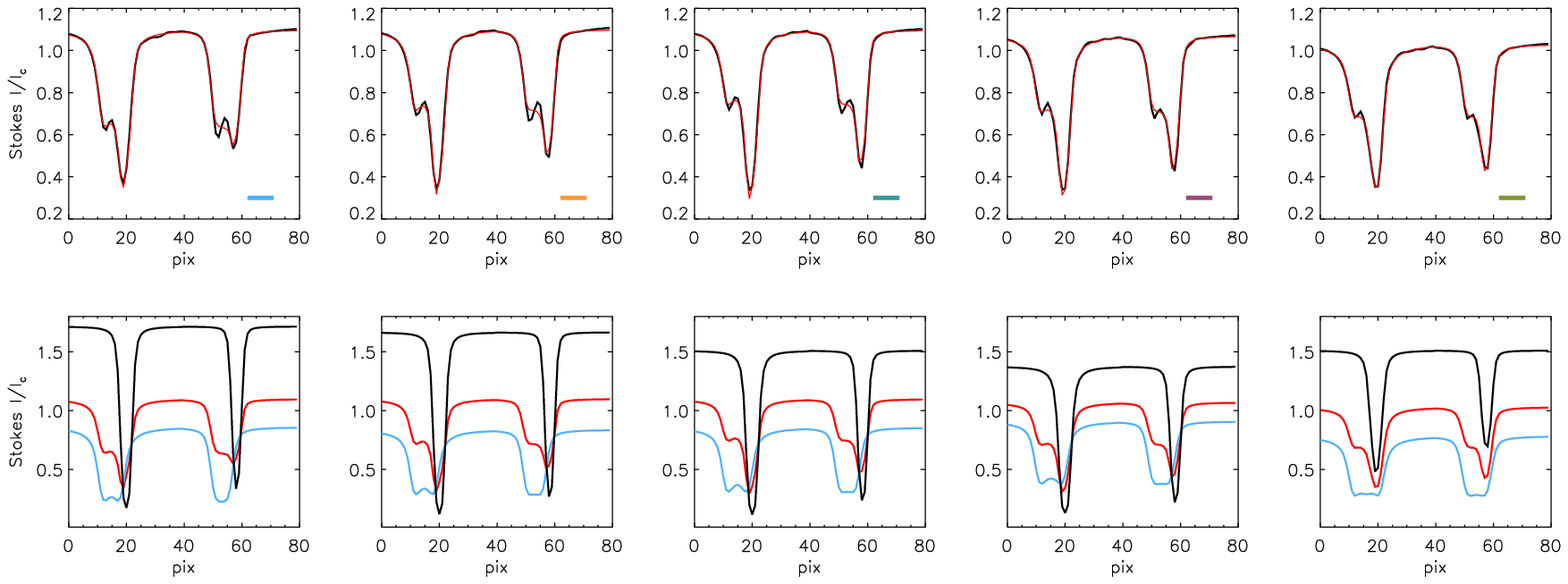}
\includegraphics[width=16.5cm]{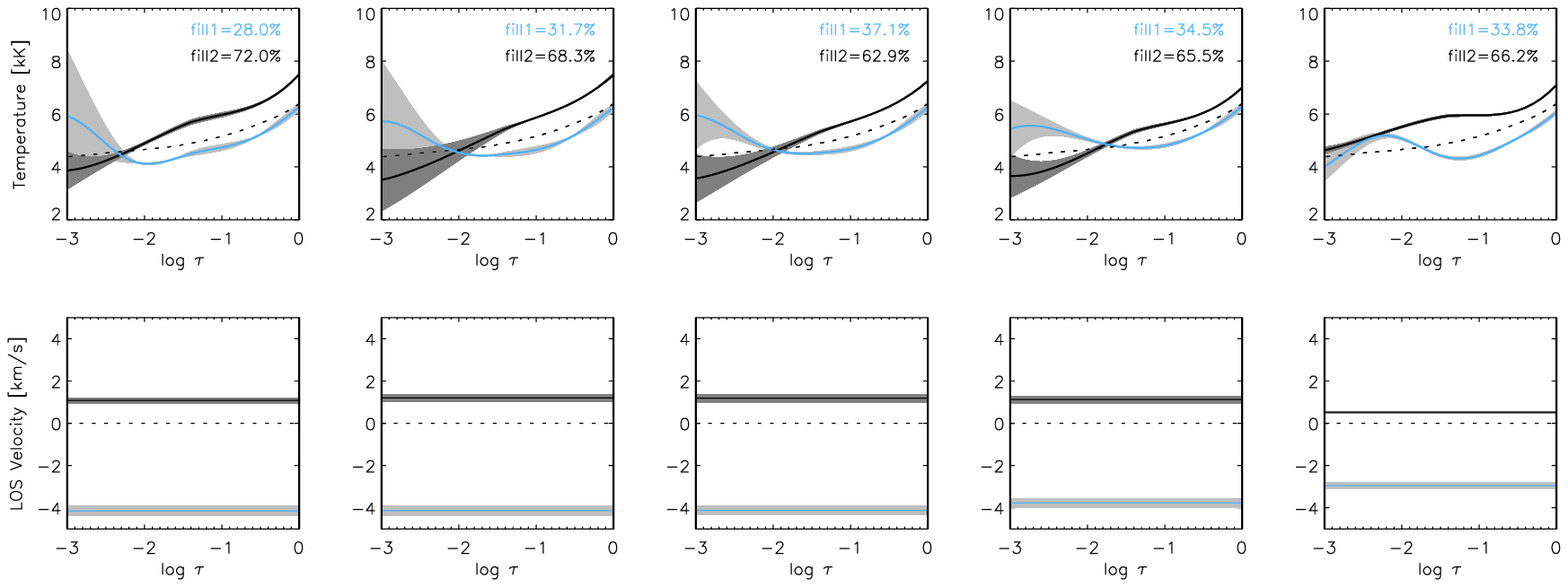}
\caption{Top row displays the original profiles in black and the resulting profiles from the inversion in red. We also indicate with coloured line at the bottom of each panel their corresponding location on Figure \ref{cont_redux}. Second row displays the two inversion components (black and blue) that compose the final inversion profile $p_f$ (red line, as in the upper panel). Third and fourth rows show the result for the temperature and LOS velocity stratification for both inversion components, following the colour code of the second row. The uncertainty of the first component (black) is plotted in dark grey while the uncertainty of the second component (blue) is displayed in light grey. The filling factor of both components is presented in the third row, at the top of each panel.}
\label{inverpix}
\end{figure*}

The maximum number of nodes for the physical parameters of the first component are five for T($\tau$)\footnote{The parameter $\tau$ refers to the optical depth evaluated at a wavelength where there are no spectral lines (continuum). In our case this wavelength is 5000 \AA.} and, one for the LOS component of the velocity V$_{\rm LOS}$($\tau$). Concerning to the second component, we use seven nodes for the temperature T($\tau$) stratification, and one node for the LOS velocity stratification. We also invert the microturbulence of the second component with one node due to the difficulty of obtaining good fits of the satellite lobe. The necessity of microturbulence implies the existence of unresolved LOS motions, something feasible if we take into account that the observed map points to a region with heliocentric angle values lower than 0.45. Finally, the macroturbulence is null and not inverted. 

The process to fit the profile implies generating two components. Thus, the final profile $p_f$ that will fit the observed Stokes parameter is a linear combination of two profiles weighted by a filling factor. This filling factor takes into account how much fraction of the observed pixel is occupied by each one of the two components. Therefore, the retrieved inverted profile can be written as 
\begin{equation}
p_f=ff\times p_1+(1-ff)\times p_2
\end{equation}
where $ff$ is the filling factor while $p_1$ and $p_2$ are the Stokes profiles from the two inverted components. It is interesting to note that the spatial resolution of $Hinode/SP$ guarantees that we can use a single magnetic component to infer the information in most of the observed scenarios. However, there are some occasions where the complex shape of the Stokes profile indicates that the structures inside the pixel are unsolved. For example, sunspot penumbral pixels are prone locations to use two different components  \citep{delToroIniesta2001,BellotRubio2004,Borrero2004}. In the present case, the complex Stokes $I$ profile shape, with clearly two distinct components, also indicates that the best way to face the problem is to use two components. Thus, the total number of free parameters, including the filling factor, is 16.

The inversion process is carried out convolving the synthetic profiles, at each iteration, with the spectral transmission profile of $Hinode/SP$ \citep{Lites2013}. Additionally, given that the inferred physical parameters could be reliant on the initial atmosphere, we minimize this effect by inverting each individual pixel with 100 different initial atmospheric models. These initial random atmospheric models were created as in \cite{QuinteroNoda2015}. 

Finally, we set the zero velocity reference using the velocity inferred from the inversion of the mean Stokes $I$ profile. This mean profile was computed in four different heliocentric regions that extend in the full horizontal range of Figure \ref{cont} and only on a small part of the vertical range of the same figure where the heliocentric angle moderately changes. Therefore, all the velocity results we are going to study in following sections are defined respect to the velocity of the mean Stokes $I$ profile.

\subsection{Inversion results}\label{res}

We show in the first row of Figure \ref{inverpix} the results of the inversion of the Stokes $I$ profiles examined in Section \ref{profile}. The spatially deconvolved observed profiles are depicted in black colour while the inverted profiles are presented in red. We can see that the fitting results are fairly good, mainly in the rightmost panels. If we examine the second row, we can find the two different components the inversion code used to generate the fitted profile. We chose black colour for the slightly redshifted component while blue depicts the blueshifted component. We also included in red colour the resulting Stokes $I$ profile the code uses to compare with the observed profile, following the same colour code used in the first row. The behaviour of the different pixels is almost the same for all panels. The redshifted component is very narrow and displays a large continuum intensity value. On the other hand, the blueshifted component shows low continuum values but the depth of the line, i.e. the difference between the continuum and line core intensities, is very small, pointing to the presence of some kind of heating at the line core formation region. Additionally, this component is highly blueshifted, being the line core in the same wavelength position of the satellite lobe found in the observed profile.

\begin{figure*}
\centering
%\hspace{-0.9cm}
\includegraphics[width=12.5cm]{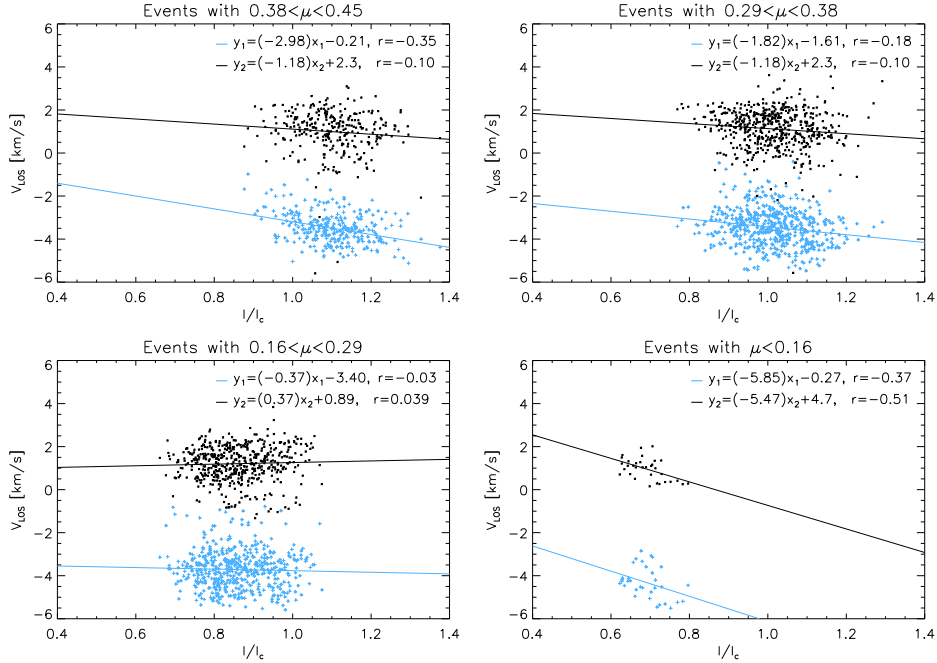}
\caption{LOS velocity values versus continuum intensity. We plot in black colour the results of the slightly redshifted component while blue colour displays the results of the blueshifted component. We indicate on the top of each panel the results of a linear fit of the values as the correlation between them. We separate the pixels located at different heliocentric angles in different panels because the continuum intensity values change with $\mu$.}
\label{vel_stat}
\end{figure*}

Third row of Figure \ref{inverpix} shows the results for the temperature stratification for both components. We also added the uncertainty provided by the inversion code. Large grey areas indicate the heights where the response of the Stokes profiles to changes in the atmospheric parameters is low, consequently, the error in the physical parameter is substantial. Additionally, we added a reference dashed line that corresponds to the Harvard Smithsonian Reference Atmosphere \citep[HSRA,][]{Gingerich1971}. If we focus first in the temperature corresponding to the redshifted component, black line, we found hot values at the continuum level (around $log$ $\tau=0$) that produce the enhanced continuum level we observed in the second row. The temperature stratification changes with height being cooler than the reference atmosphere for the rest of the atmosphere. The reason for that is probably the necessity of generating a very narrow and deep Stokes $I$ profile. On the other hand, the blue line, blueshifted component, exhibits an opposite behaviour. The temperature stratification is cool at the low photosphere, generating low continuum intensity values. Then, the temperature stratification gradient changes and the atmosphere becomes very hot in the upper layers, with values of more than 2000 K larger than the reference atmosphere. We believe that this temperature enhancement at upper photospheric layers is the responsible of the Stokes~$I$ low line depth mentioned before. Moreover, the enhancement is high enough to produce a line core in emission for the Fe~{\sc i} 6301 \AA \ line, which has a higher height of formation. Moreover, the optical depth where the change of gradient takes place, monotonically changes from the centre of the granule (leftmost panel) to the edge of the granule (rightmost panel), happening at lower heights as we move to the periphery of the granular structure.

\begin{table*}
  \caption{Mean LOS velocity and filling factor ($ff$) values for the two inverted components.}
  \label{table}
  \begin{tabular}{lccc}
	\hline
	   $\mu$    &  $C_1$ $<V_{LOS}>$ [km/s] & $C_2$ $<V_{LOS}>$ [km/s] & $C_2$ $<ff>$  \\
	\hline
	0.38$<\mu<$0.45       &  -3.32 $\pm$ 0.74  &   1.14 $\pm$ 1.04     & 69.5 $\pm$ 12 \% \\
	0.29$<\mu<$0.38       &  -3.29 $\pm$ 0.85  &   1.27 $\pm$ 0.98     & 66.3 $\pm$ 13 \% \\
	0.16$<\mu<$0.29       &  -3.53 $\pm$ 0.87  &   1.39 $\pm$ 0.80     & 63.7 $\pm$ 11 \% \\
	\qquad \ \ $\mu<$0.16 &  -4.16 $\pm$ 0.74  &   1.12 $\pm$ 0.50     & 65.3 $\pm$ 10  \% \\
	\hline
  \end{tabular}
\end{table*}

\begin{figure*}
\centering
%\hspace{-0.9cm}
\includegraphics[width=12.5cm]{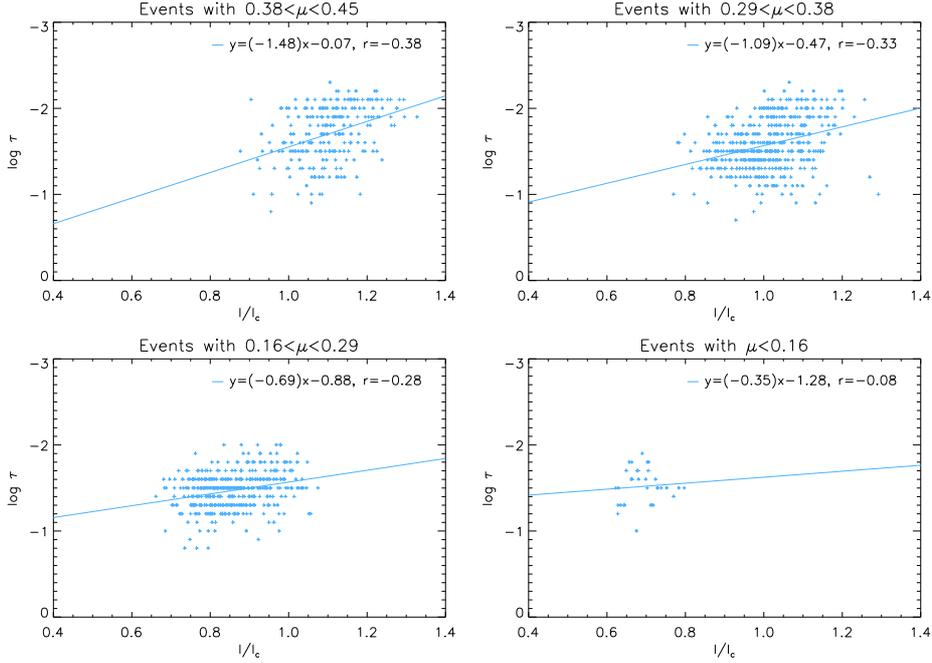}
\caption{Each panel represents the location of the temperature minimum (in $log$ $\tau$ units) mentioned in Section \ref{res} versus the continuum intensity. As in Figure \ref{vel_stat}, we calculate a linear fit of the obtained values as the correlation between them. We excluded the redshifted component of this plot because is not relevant. Additionally, we also separate the results in different panels depending on the heliocentric angle of the studied pixel.}
\label{temp_stat}
\end{figure*}

Regarding the filling factor of the inverted components, see top part of each panel of third row, we can see that the blueshifted component occupies a small fragment of the pixel, with values around 30 per cent of the total area. These values are coherent with the fact that the blueshifted component represent a satellite profile which importance on the observed Stokes $I$ profile is lower than that of the central component. 

Bottom row shows the results of the LOS velocity stratifications. The redshifted component, black colour, displays redshifted values lower than 1.5~km s$^{-1}$ for all pixels. On the other hand, the blue component displays blueshifted velocities that start with 4.2~km s$^{-1}$ in the centre of the granule and then decreases until 2.5~km s$^{-1}$ when we reach the periphery of the granular structure. Finally, the microturbulence of the blueshifted component, not shown in this figure, displays values between 1.5 and 2.0~km s$^{-1}$. The reason for these values is to produce a sufficiently broad profile and, at the same time, compensating the line core intensity to improve the profile fit. 

As additional comment, the inversion of the satellite lobe in the first and second panel, first row starting from the left, is not perfect. We performed additional tests aiming to reproduce these two profiles and we only obtain good fits when we provide 7 nodes to the atmospheric LOS velocity stratification of the second component. In this case, the resulting atmosphere displays large variations along the line of sight that were difficult to understand and sometimes erratic between the selected pixels. Therefore, we stay with the previous configuration for the present and following sections, easier to interpret, although it does not provide perfect fits.

\section{Statistical analysis}

We inverted all the pixels that show a satellite lobe in the blue wing of the Stokes $I$ profile, red points on Figure \ref{cont}. The total number of pixels is 1415 and we followed the same inversion configuration described in Section \ref{config}. We aimed to verify if the relation between the continuum intensity and the temperature and LOS velocity stratifications that we found in the previous section is generally reproduced. We have to bear in mind, that we are going to analyse pixels from different heliocentric angle regions and different events, therefore, a large deviation in the results is expected. Additionally, as the continuum intensity level decreases with the heliocentric angle, we need to separate the found events as function of their heliocentric angle to properly account their evolution with the continuum intensity level.

We plot in Figure \ref{vel_stat} the results for the LOS velocity. A large variation can be found for the satellite component, blue colour, although it seems that the LOS velocity decreases as we move to the edge of the granule, i.e. to lower continuum intensity values. We computed a linear fit of the scatter values to estimate the previous assumption. We found that, in most of the cases, the slope of the linear fit has negative values pointing out that the blueshifted velocity decreases as the continuum intensity becomes smaller. However, the correlation between values is very low, being always lower than 0.35, indicating that there are a large number of pixels that strongly departure from the fitted line. On the other hand, the redshifted component, black colour, shows low positive values, redshifted velocities, and a random dependence with the continuum intensity. 

We also compute the mean LOS velocity and filling factor values, as their standard deviation, for both components at different heliocentric angles. The obtained values are shown in Table \ref{table}. We can see that the mean velocity and filling factor barely change with different heliocentric angles, being close to -3.5 km s$^{-1}$ for the blueshifted component and around +1 km s$^{-1}$ for the redshifted component. The same happen with the filling factor between both components, being around 65-70 per cent for the redshifted component for all heliocentric angles. These results are in agreement with the ones found in Section \ref{res}, where the satellite lobe is generated by a component that occupies a small fraction of the total pixel area and has blueshifted velocities lower than 5 km s$^{-1}$.

We mentioned on Section \ref{res} the existence of a change on the temperature gradient of the blueshifted component that translates into a hot region at upper layers. We retrieved the position of the change on the temperature gradient, i.e. the temperature minimum in the range of optical depths showed in Figure \ref{inverpix}, to study its dependence with the continuum intensity. We show the results in Figure \ref{temp_stat}. All panels display linear fits with negative values indicating that the change of gradient, and the consequent increase of temperature at high layers, is taking place at lower heights as we move to the edge of the granule, i.e. to lower continuum intensity values. However, as happen before, the correlation between values is low and, thus, we should bear in mind that there are many pixels that apparently deviate from this behaviour.

\section{Summary and discussion}

We have examined a spectropolarimetric observation of the north solar pole. Taking advantage of the spatial deconvolution technique we studied the Stokes $I$ profiles that show a satellite lobe in the blue wing. We established that their density is similar to what other authors have found, see \cite{BellotRubio2009}, and smaller than the predicted values from simulations \cite{Stein1998}. In addition, we determined that the spectral position of the satellite lobe evolves with the spatial location on the granule, being located close to the Stokes $I$ line core when the pixel belongs to the centre of the granular structure and appearing at the blue wing of Stokes $I$ when the pixel belongs to the edge of the granule. These findings are also in agreement with \cite{BellotRubio2009}. Later, we made use of inversion techniques to comprehend the atmospheric physical parameters that describe the satellite lobe as its evolution. As we removed the stray light contamination produced by the spatial point spread function of the telescope, we carried out the inversion without any stray light component. However, the complexity of the problem demands using two inversion components, one to reproduce the presence of the satellite lobe and the other to generate the almost non-shifted absorption component. As the polarimetric signals on the selected pixels were low, most of the times at the noise level, we only inverted the Stokes $I$ parameter using two non-magnetic components. We obtained good fits in most of the inverted pixels although the fit often was more accurate for the Fe~{\sc i} 6301.5 \AA \ line. We believe that the reason behind is the difference between the height of formation of both lines. It is possible that the fit will improve if we include gradients along the LOS velocity stratification. However, we perform several tests in this regard and we were unable to consistently improve the fits.

The inversion results reveal the coexistence of a blueshifted component that is hot at upper layers with an almost slightly redshifted component that is cool at upper layers. The presence of the redshifted component in the pixel is considerably larger occupying around 65-70 per cent of its total area, while the rest of the pixel is filled with the satellite component. We found that the LOS velocity of the latter component monotonically changes from the centre to the edge of the granule, reducing its maximum value as we approach the limit of the granular structure. Moreover, we found a change on the temperature gradient of this component that also shifts its location from upper to lower layers as we move to the borders of the granule. After these findings, we proceeded to examine all the selected pixels that fulfil our threshold criterion. They belong to different cases that are located at distinct heliocentric angles. We found that the heliocentric angle barely changes the results and that they are consistent with the findings obtained for the case examined in detail.

However, in spite of our results, there are some aspects that we still need to study to completely understand these horizontal flows. The first one is their evolution. We cannot infer temporal information from the map used in this work. Hence, we plan to analyse time series observations to address this question. The second aspect is the physical origin of these horizontal flows. In that sense, we plan to make use of numerical simulations to synthesise the Stokes profiles from quiet Sun areas at different heliocentric angles looking for profiles that show spectral features similar to the ones examined in this work. Thus, we expect to increase the knowledge about these events in future studies.

\section*{Acknowledgements}
We thank B. Ruiz Cobo his helpful comments and suggestions after carefully reading the manuscript. \textit{Hinode} is a Japanese mission developed and launched by ISAS/JAXA, collaborating with NAOJ as a domestic partner, NASA and STFC (UK) as international partners. Scientific operation of the Hinode mission is conducted by the Hinode science team organized at ISAS/JAXA. This team mainly consists of scientists from institutes in the partner countries. Support for the post-launch operation is provided by JAXA and NAOJ (Japan), STFC (U.K.), NASA, ESA, and NSC (Norway).

\bibliographystyle{mnras} % style apj.bst
\bibliography{hor} % your references Yourfile.bib

% Don't change these lines
\bsp	% typesetting comment
\label{lastpage}
\end{document}